\input{aipcheck}

\documentclass[
    ,final            
  ]
  {aipproc}

\layoutstyle{6x9}

\begin{document}

\title{Flux tube counting or Casimir scaling}

\author{Sedigheh Deldar}{
  address={Department of Physics, University of Tehran, Iran}
}

\author{Shahnoosh Rafibakhsh}{
  address={Department of Physics, University of Tehran, Iran}
}

\begin{abstract}
QCD confines quarks in all representations. From lattice calculations 
and fat center vortices model, we discuss that the coefficient 
of the linear term in the potential is proportional to
both casimir scaling and the number of fundamental strings.
\end{abstract}

\maketitle



Explaining the confinement of quarks in QCD is still a challenging problem.
Even though perturbative techniques describe very well the behavior of quarks
at short distances, the mechanism of confinement which prevents quarks and
gluons to be found free, is still a puzzle. Many lattice calculations
confirm the confinement of quarks in the fundamental and higher 
representations at intermediate distances, but phenomenological models have 
not been fully successful in describing the infrared
behavior of quarks and gluons. One of the features of the
confinement is that the potential between quarks increases by distance.
Recent numerical calculations \cite{Deld2000,Bali2000} show that the string tension,
the coefficient of the linear term in the potential, is representation
dependent and proportional to the eigenvalue of the quadratic casimir operator
of the representation. This proportionality is called ``Casimir scaling".
The proportionality of the potential with casimir operator is expected for
short distances where the force between quarks can be described by one gluon
exchange and the coupling is proportional to the quadratic
casimir operator. But this behavior is still not understood for intermediate
distances. On the other hand, some people believe 
the string tension between two quarks of a representation can be 
obtained by multiplying the fundamental string tension times the number of 
fundamental flux tubes embedded into that representation.
The fundamental (elementary) string is a string which connects a fundamental heavy quark
with an anti-quark.
In this paper, we take a closer look at lattice results and explain that string 
tensions may also agree with the flux tube counting idea.  
We also discuss results obtained for potentials between 
quarks from the fat center vortices model for SU(2), SU(3) and especially our
recent calculations for SU(4) and show that string tensions
are in agreement with both casimir scaling and flux tube counting but they
agree better with flux tube counting especially for
larger gauge groups. 

Although lattice calculations show a qualitative agreement between string 
tensions and casimir operators,
looking more carefully at lattice results, one can interpret string ratios with
the idea of flux tube counting. Cross signs on the plot on the left hand side 
of figure 1 show ratios of string tensions of SU(3) sources of various 
representations to that of the fundamental
representation obtained from lattice calculations \cite{Deld2000}.
In this paper, we sometimes call sources in higher representations as ``quarks".
Circles indicate casimir ratios and diamonds show the number of fundamental
strings in each representation. As claimed by lattice people, ratios of
string tensions are proportional to casimir operators. On the other hand,
the plot shows that there is a rough agreement with the number of
fundamental tubes as well. A. Armoni {\it et al.} \cite{shif} have explained 
why ratios from the
lattice are larger than the number of fundamental fluxes. If we define the
potential between two quarks as the potential between fundamental strings, then, when
strings are very far from each other, they do not have any interaction 
and the string tension is equal to the string tension of quarks
in the fundamental representation times the number of strings. On the other hand,
if we put quarks in an appropriate distance, the elementary strings attract
each other and this attraction reduces the string tension to some
number less than the potential of elementary strings. If quarks get even
closer, an overlap between strings happens. Therefore, they repel each
other and the string tension between two sources decreases
such that it gets larger than the potential between fundamental tubes.
As Armoni {\it et al.} have discussed, the typical length/thickness ratio of the
fundamental string of lattice calculations is not large enough. Thus,
an overlap between fundamental strings may exist which leads to a repulsion
and therefore makes string ratios larger than the number of fundamental
fluxes. 

\vspace{3.pc}
\begin{figure}[h]
  {\includegraphics[height=.24\textheight]{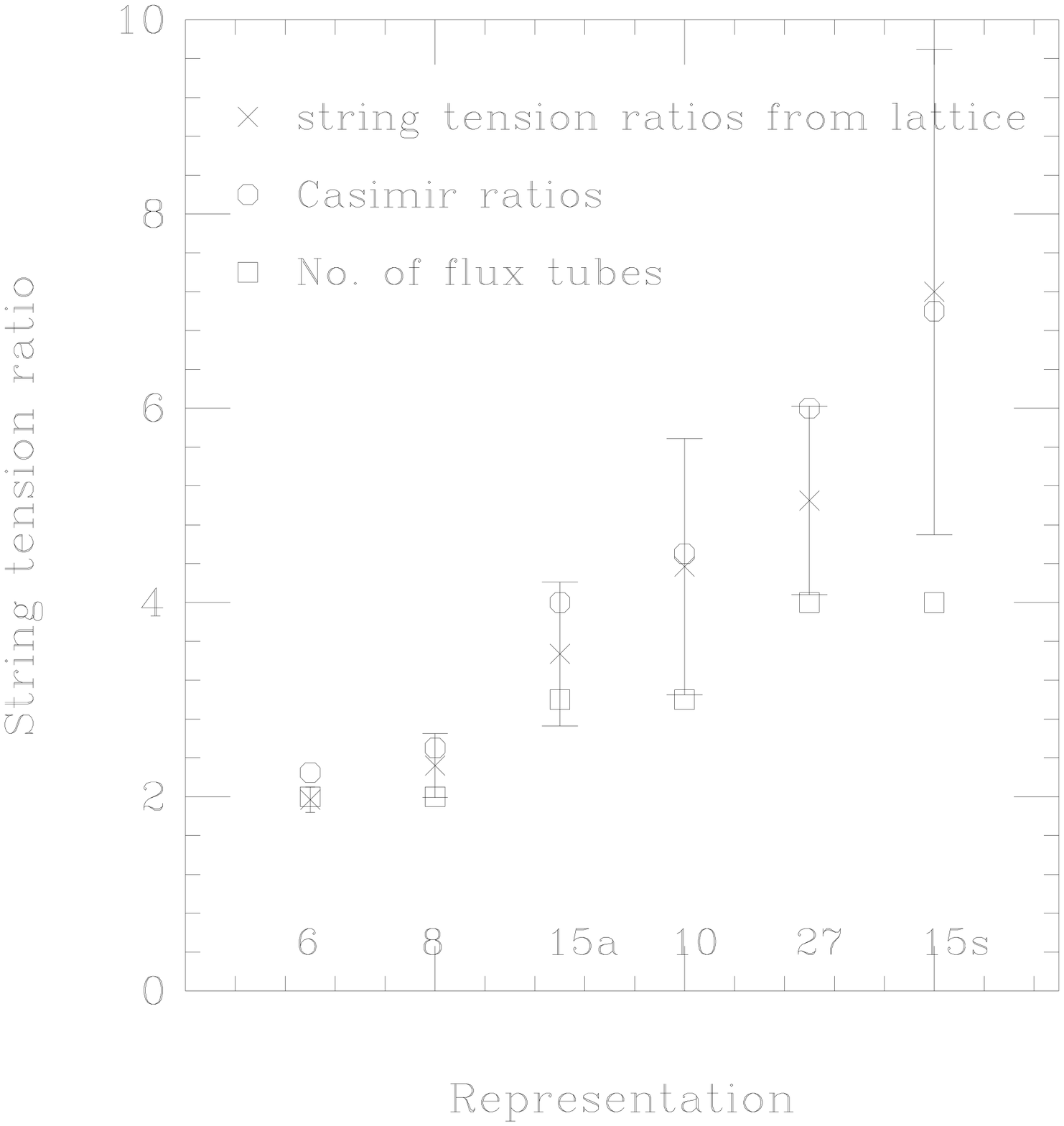}}
  \hspace{2pc}
  {\includegraphics[height=.23\textheight]{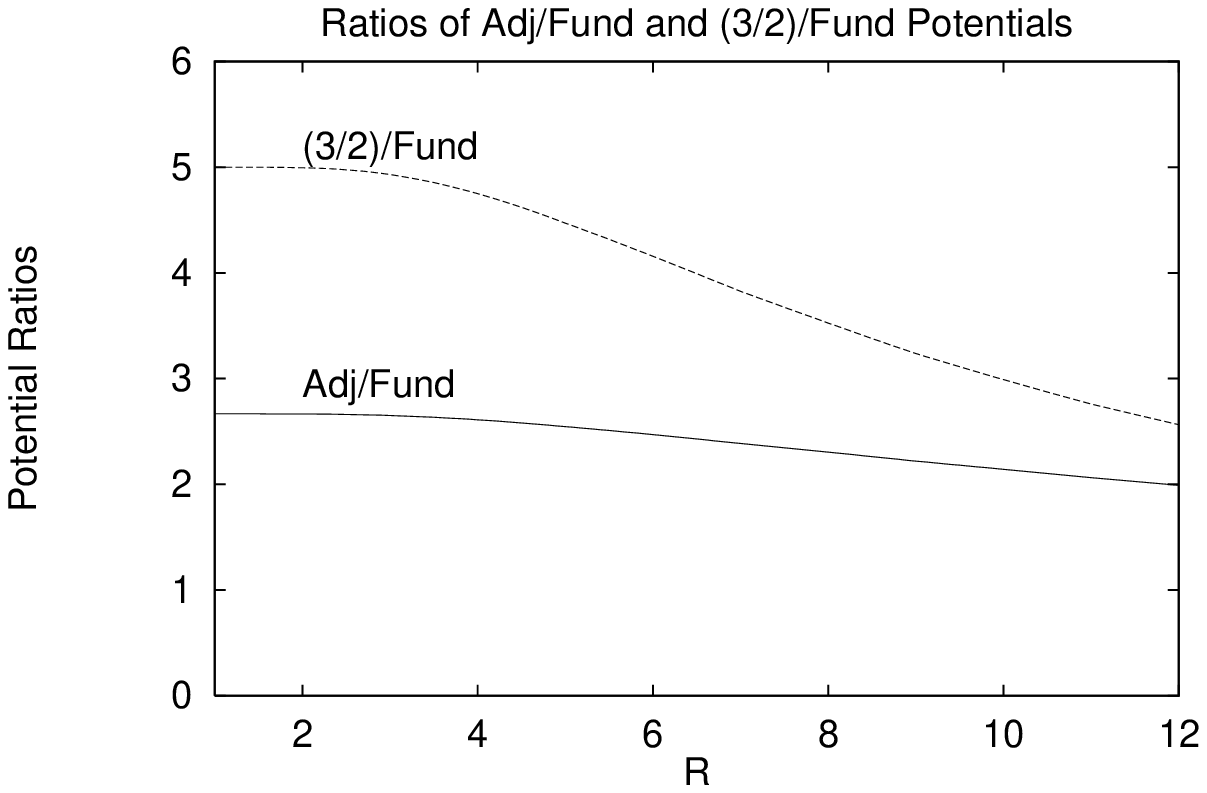}}
\vspace{-10pc}
  \caption{The left plot shows ratios of string tensions of SU(3) quarks of
different representations to the string tension of quarks in the fundamental 
representation. Considering the lattice data errors, a good agreement with both
casimir scaling and flux tube counting is observed. Potential ratios between
quark charges 
of $j=1$ (adjoint) and $j=3/2$ to that of the fundamental quark ($j=1/2$)
are shown in the right. Ratios start up at corresponding casimirs which are 
$8/3$ and $5$ for $j=1$ and $j=3/2$, respectively.}
\end{figure}

Now, we discuss results obtained from the fat center vortices model for SU(2), SU(3)
and SU(4) gauge groups. The center vortices model is a phenomenological model
which claims that the QCD vacuum is a condensate of color
magnetic vortices. 
These vortices are responsible for confining color charges.
For intermediate distances, center vortex model has predicted the confinement 
of quarks in the fundamental representation. M. Faber {\it et al.}
\cite{Faber} have observed the confinement of quarks for all representations
by making the vortices thick enough. 
The plot on the right hand side of figure 1 shows ratios of string tensions for
SU(2) sources \cite{Faber}. Potentials for quark charges in the $j=1/2$, $1$,
$3/2$ are calculated and potentials ratios of sources with the $j=1$ 
(adjoint) and $j=3/2$ to that of the fundamental quark ($j=1/2$) are plotted.
As indicated in the figure, ratios start up at of casimir ratios which are 
$8/3$ and $5$ for $j=1$ and $j=3/2$, respectively. Figure 2 shows potential
ratios of quarks in various representations to that of the fundamental one,
for SU(3) \cite{DelJhep} and SU(4) \cite{Delshah} gauge groups. Again,
ratios start up roughly at ratios of corresponding casimirs
but change so that at some region, which is different for each representation,
get close to the number of fundamental strings of that representation.
The agreement with flux tube counting in the most linear part of the potential 
is better for SU(4) than SU(3) \cite{IPM}. This is in agreement with
reference \cite{Bali} which claims that by increasing the number of gauge
groups, the interaction between fundamental strings decreases and the total
string tension would be the fundamental string tension times the number of
strings.

\vspace{3pc}
\begin{figure}[h0]
  {\includegraphics[height=.29\textheight]{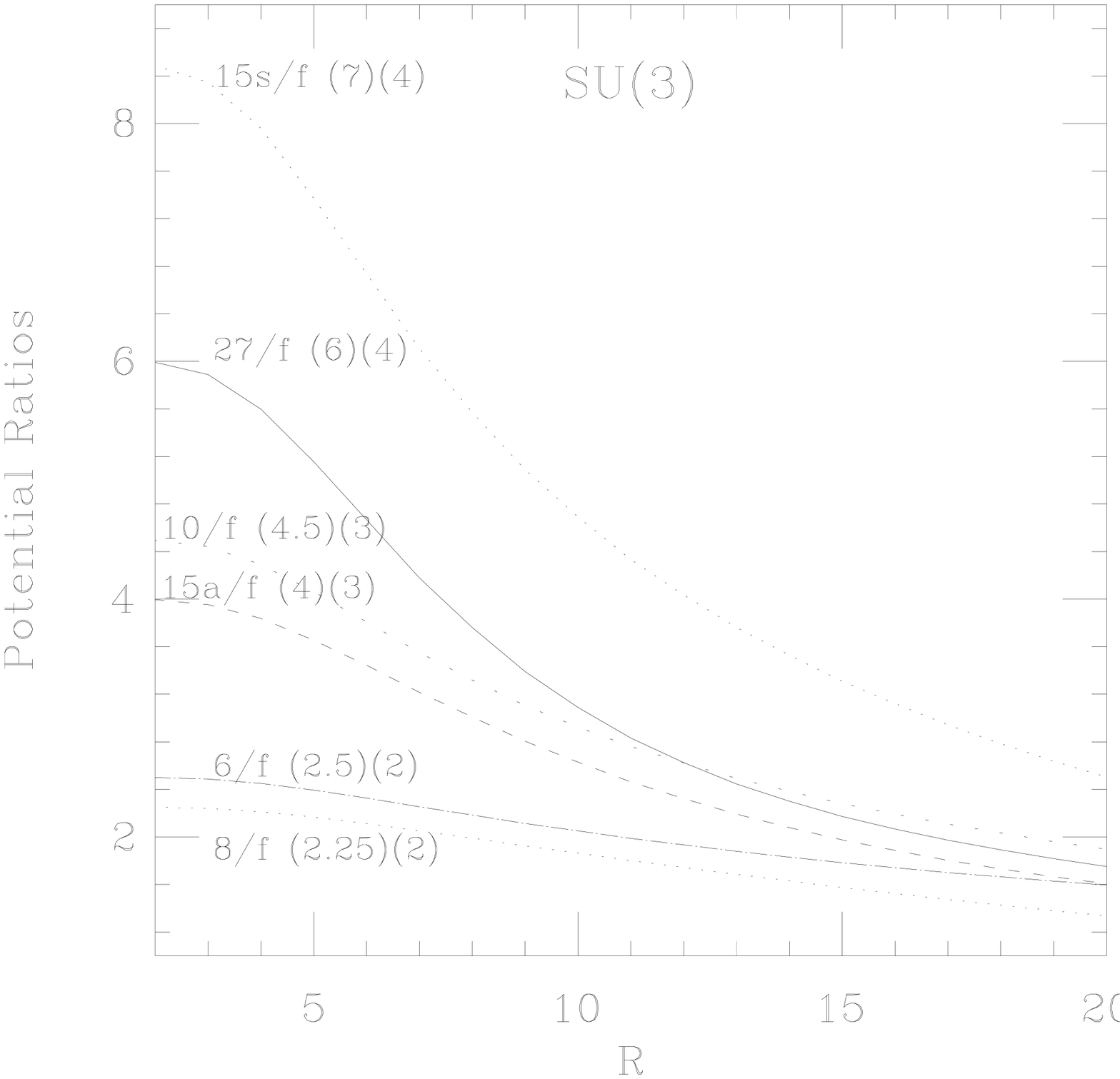}}
  \hspace{2pc}
  {\includegraphics[height=.29\textheight]{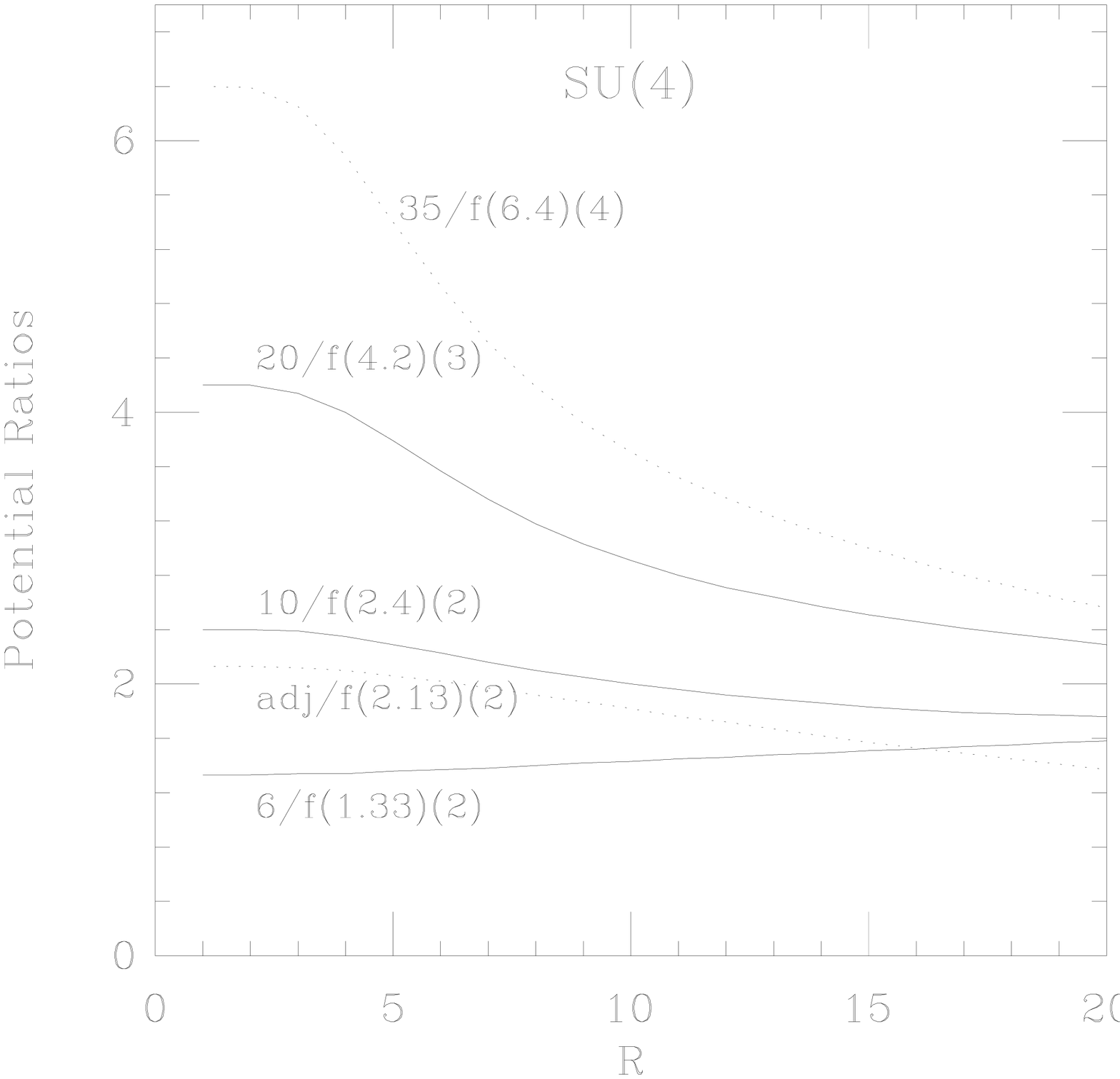}}
  \caption{Ratios of potentials between quarks of higher representations to
that of the fundamental one, for SU(3) and SU(4) gauge groups. Casimir ratios and the number of fundamental
strings are shown in the first and second parentheses, respectively.
Ratios start at corresponding casimirs but get close to 
the number of fundamental strings.
}
\end{figure}
We conclude that both lattice calculations and the fat center vortices model predict
a linear regime for the potential between quarks of the fundamental and
higher representations. The string tension in that region is proportional 
to both casimir scaling and the number of fundamental tubes. The proportionality
with the number of fundamental tubes seems to be better from fat center vortices
model especially for SU(4) gauge group.

We would like to thank the research council of the University of Tehran for
supporting this work.

\end{document}